**Driving atomic structures of molecules, crystals, and complex systems with local similarity kernels**

*Ziheng Lu[1*], Wenlei Shi[1*], Lixin Sun[2], Haiguang Liu[1], Tie-Yan Liu[1]*

[1] *Microsoft Research Asia, No. 5 Dan Ling Street, Haidian District, Beijing 100080, China*

[2] *Microsoft Research Cambridge, 21 Station Road, Cambridge CB1 2FB, United Kingdom*

[*]*Correspondence to:*

Ziheng Lu <zihenglu@microsoft.com> and Wenlei Shi <wenlei.shi@microsoft.com>

**Abstract**

Accessing structures of molecules, crystals, and complex interfaces with atomic level details is vital to the understanding and engineering of materials, chemical reactions, and biochemical processes. Currently, determination of accurate atomic positions heavily relies on advanced experimental techniques that are difficult to access or quantum chemical calculations that are computationally intensive. We describe an efficient data-driven LOcal SImilarity Kernel Optimization (LOSIKO) approach to obtain atomic structures by matching embedded local atomic environments with that in databases followed by maximizing their similarity measures. We show that LOSIKO solely leverages on geometric data and can incorporate quantum chemical databases constructed under different approximations. By including known stable entries, chemically informed atomic structures of organic molecules, inorganic solids, defects, and complex interfaces can be obtained, with similar accuracy compared to the state-of-the-art quantum chemical approaches. In addition, we show that by carefully curating the databases, it is possible to obtain structures with bias towards target material features for inverse design.

**Keywords:** atomic structures, atomic environments, similarity kernel, structure optimization, inverse design

**Introduction**

Access to atomic-level structural information of materials, interfaces, and processes is crucial to the understanding of the physical world, laying the foundation of modern chemical/materials science and engineering. [1-4] Conventionally, such a task has been achieved with experimental characterization techniques such as X-ray diffraction, transmission electron microscopy, and scanning tunneling microscopy. [5, 6] Depending on the system of interest, specialized techniques may be needed. For example, neutron diffraction is used to characterize the position of light-weight elements such as lithium. [7] Nuclear magnetic resonance and electron energy loss spectroscopy are used to probe the local structural information. [8] While these techniques have significantly advanced our understanding on the arrangement of atoms, many of them heavily rely on the infrastructure, posing limits on the throughput and broad applications.

During recent decades, computational methods provide an alternative/complement to experiments. Such methods elucidate critical atomic structures of molecules and materials by sampling from the Boltzmann distribution that are determined by the potential energy surface (PES):

$$E = E(\{x_i\}_{i=0}^n) \qquad (1)$$

where $x_i$ is the Cartesian coordinate of the $i$th atom in the system and $E$ is the potential energy arising from a particular set of $x_i$. The gradient of energy with respect to $\{x_i\}_{i=0}^n$, can be used to drive the configuration to the regions of interest. This gradient can aid minimization algorithm to find the structures of local or global energy minima, which correspond to the dominating ground state structures at low temperatures. It can also be used in modeling kinetics at finite temperatures with molecular dynamics and Monte Carlo sampling.

One way to compute $E$ is to use quantum chemical approaches such as density functional theory (DFT) or post-Hartree-Fock methods to numerically solving the electronic structure, as illustrated in **Figure 1(a)**. Obtaining atomic structures with quantum chemical approaches has achieved great success in the area of batteries, superconductors, catalysts, chemical reactions, and biological molecules. [9-11] Despite so, energy evaluation based on quantum mechanics suffers from intensive computational demands. Considering the $O(N_e^3)$ to $O(N_e^4)$ complexity of current quantum chemical methods such as DFT and its hybrid cousins ($N_e$ being the number of electrons), typical system sizes are constrained to hundreds of atoms with a spatial extent of less than ten nanometers. This severely constrains the applicability of these methods to structures with large number of atoms. For example, medium sized protein molecules can contain a few hundred amino acids, i.e., thousands of atoms. Considering the aqueous environment of these biomolecules, explicit representation of water molecules will increase the system size to 100k, which is intractable for quantum mechanics calculations.

One solution to reduce the computational cost is to describe the interaction between atoms using classical force fields without explicitly considering electrons. Instead, only the positions of atom nuclei are used in computing the PES, as shown in **Figure 1(b)**. An example classical force field for biomolecule includes function of internal coordinates, e.g., bond lengths, bond angles, and dihedrals. Thanks to the low cost $O(N)$ complexity of force fields ($N$ being the number of atoms), it is possible to simulate large systems composed of millions of atoms. However, classical force fields usually require a predefined topology of molecules, so that chemical processes such as bond formation/break cannot be correctly simulated. This drawback severely limits the applications of such classical models, especially in modeling chemical processes.

Recently, machine learning potentials (MLPs) have been developed by fitting the PES using a machine leaning model such as a Gaussian process regression or neural networks. [13, 14] MLPs have been proven to have both the accuracy of quantum mechanical methods and the linear O(N) scaling of classical force fields. [15] However, despite being more expressive than classical force fields, MLPs still suffers from transferability issues where the MLP trained on one system will fail for a different system. Moreover, despite a few attempts in optimizing MLP for experimental observables, most MLP are trained on structural data with potential energy or force labels. This prevents the use of unlabeled data and data generated using different quantum mechanical methods, thereby hindering the use of structural data accumulated over the years.

A typical example is shown in **Table 1** where the direct energy output for a C2/m $LiCoO_2$ crystal, a widely used cathode material used in lithium-ion batteries [29], has a large variance of ~2.9 eV atom$^{-1}$. In fact, even if one considers the formation energy, the adoption of different correction schemes, e.g., Hubbard U correction, makes the values not directly comparable. [16, 17] In contrast, the structures from these databases are similar. Their difference in density is on the order of 0.1 g cm$^{-3}$, < 2% relative error.

Similar issues can also be found in organic molecules. As shown in **Figure 2(b) and (c)**, the energy variance of QM9 dataset evaluated using different exchange-correlation functionals are on the order of 0.5 meV atom$^{-1}$ while the structural local minimum is relatively close with a mean squared deviation of ~0.2 Å. [18] The vast differences of energies for highly similar structures call for a geometry-based optimization method, so that on the computed energy reflect the geometry of structures. This optimization method should also be capable of utilizing the structure datasets from different databases, including those structures with energy values.

Interestingly, despite the vast structural space, the local atomic environments in stable materials (or low energy materials) are limited. **Figure 2(a)-(f)** show the atomic environments from randomly sampled carbon crystals and from low energy crystals. [19, 20] To visualize the coverage of the local structural space by different crystals, we carried out principal component decomposition on the smooth overlap of atomic position (SOAP) fingerprints. [21, 22] As shown in **Figure 2(f)**, the local environments of carbon atoms from relaxed crystals only cover a very small portion of those sampled randomly. In addition, the local environment of carbon in materials with relative low energies ($< 1\text{eV atom}^{-1}$) takes up even fewer points. This indicates that local structure of carbon materials in low energy states can be constructed based on a very limited number of local structural templates, as shown in **Figure 2(a)** to **(e)**. Therefore, local atomic environments provide a feasible approach to optimize atomic structures for materials in their stable states.

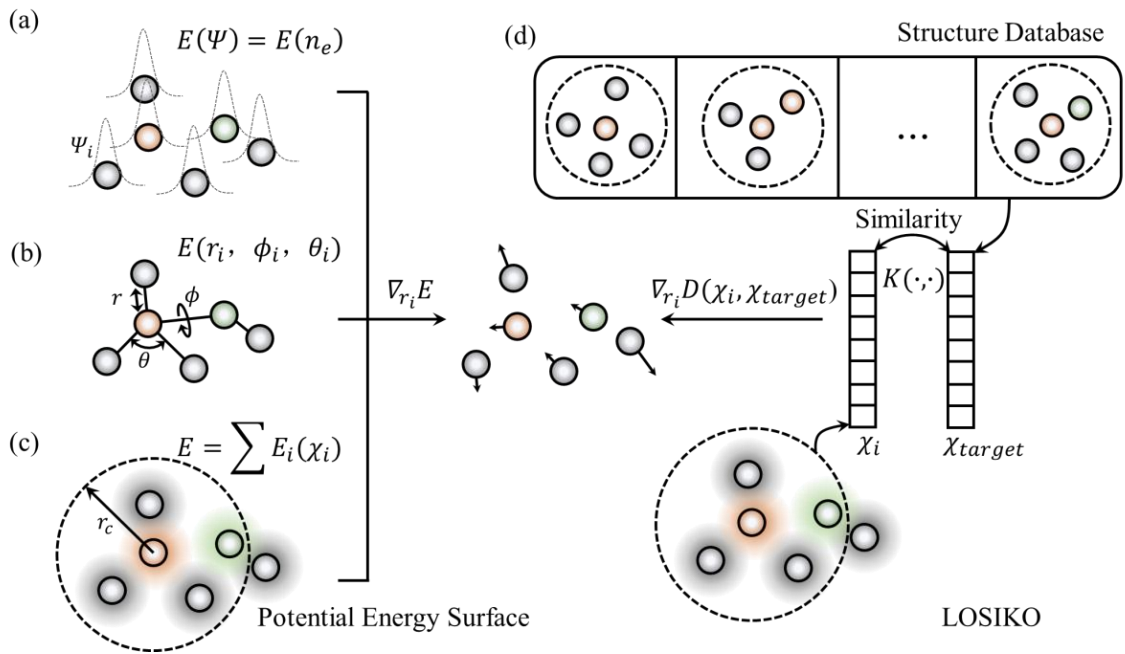

**Figure 1** *Schematics of structural optimization methods. (a) electronic structure methods, (b) classical force field, (c) machine learning force fields, and (d) data-driven LOSIKO.*

In this work, we describe an optimization scheme, named LOcal SImilarity Kernel Optimization (LOSIKO), to access chemically informed atomic structures solely based on structural features. As illustrated in **Figure 1(d)** and **Figure S1**, this method involves matching atomic environments to databases using predefined fingerprints or machine-learned embedding, followed by maximizing their similarity kernel:

$$\hat{S} = \sum_{i=0}^{n} c_i K(\chi_i, \chi_i^t) \qquad (2)$$

where $K(\cdot,\cdot)$ denotes the similarity kernel between the atomic environment $\chi_i$ of atom $i$ and its target $\chi_i^t$ in the database; $c_i$'s are the weights which are defined by the user. A larger weight means the optimization is biased towards those local structures more. In principle, any forms of local atomic environment $\chi_i$ and any form of similarity measure can be used. Within this formalism, if one confines the database to only including the low energy entries, reasonable atomic structures with low energies can likely be obtained. In addition, the database can be customized to bias the atomic structures towards desired features.

*Table 1. DFT calculated C2/m LiCoO$_2$ from different databases*

| Database | Raw Energies* eV/atom | Formation Energies* eV/atom | Density g/cm3 | Exchange-and-correlation (XC) functional | DFT Code | Additional Notes |
|---|---|---|---|---|---|---|
| Materials Project [23] | -0.687 | -2.190 | 3.63 | GGA-PBE | VASP | DFT+U |
| Javis [24] | -5.035 | -2.1832 | 3.895 | OptB88vdW | VASP | |
| AFlow [25] | -6.175 | -1.549 | 3.727 | GGA-PBE | VASP | DFT+U |
| OQMD [26] | -6.161 | -2.110 | 3.64 | GGA-PBE | VASP | |
| MaterialsGo | -8.610 | - | 3.62 | HSE | VASP | |
| MAE | **~2.9** | - | 0.1 | | | |

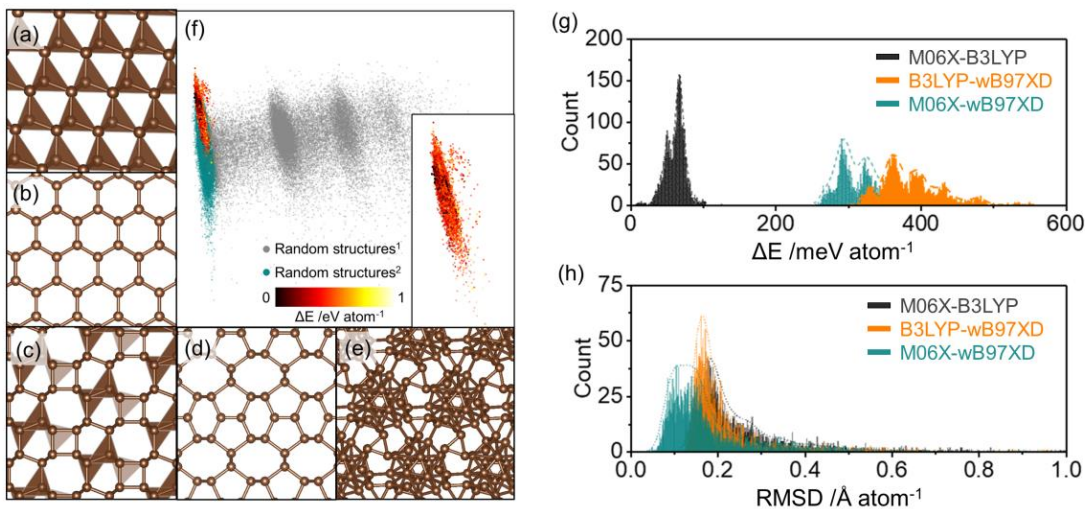

*Figure 2* *(a)-(d) Relaxed carbon structures from random structural searches. (e) An unrelaxed random carbon structure. (f) SOAP embedding of local carbon structures projected on the first two principal components. (g) The energy difference distribution between QM9 structures relaxed with different functionals. (h) The RMSD distribution between QM9 structures relaxed with different functionals.*

**Results and Discussions**

We showcase LOSIKO using a toy system, i.e., a single water molecule. The database for this case contains the SOAP embeddings of local atomic environments in a single pre-optimized $H_2O$ molecule. The loss function is defined as the sum of the $L_2$ distances, see Eq. (8), between the local environment embeddings calculated from the structure to relax and the ones in the database. Such a loss reflects how similar the local environments are between the structure and a real water molecule. Details on the computational method is discussed in the Methodology section.

By fixing two of the atoms in the molecule, two-dimensional loss surfaces (as analogue to the potential energy surface from DFT calculations) are shown in **Figure 3(a)**. When the two hydrogen atoms are fixed while the oxygen atom is allowed to move, two symmetric local minima emerge on the loss surface. Both are located 0.95 Å from the hydrogen atoms. Such minima correspond to the two chiral $H_2O$ configurations when constrained on a two-dimensional plane. Moreover, the loss monotonically decreases as the oxygen approaches the minima. This indicates optimizing the loss function will lead to the correct configuration of a water molecule no matter where the oxygen atom is placed initially. Similar results can be observed when the hydroxyl group is fixed but the other hydrogen is allowed to move. As shown in **Figure 3(b),** two local minima exist in this case, both corresponding to the correct geometry of a water molecule.

To further test the numerical stability of the optimization process, we constructed a random configuration of one oxygen atom and two hydrogen atoms, which is relaxed to a structure of the water molecule using LOSIKO. The results are shown in **Figure 3(c)**. During the optimization, the loss function value went through several stages of decrease and finally dropped to zero. Correspondingly, the randomly placed atoms are gradually driven to form a water molecule. Interestingly, we found that each stage of decreasing loss function value corresponds to one type of structural transformation. Initially, one hydrogen atom got attached to the oxygen to form a hydroxyl group (phase I. in the **Figure 3**(c)). Then, the other hydrogen was driven to bond with the oxygen as well (phase II.-III.). Finally, the H-O-H angle was adjusted to 104.5° and the optimization was finalized (phase IV.-V.). Such behavior may be related to the choice of the embedding of local atomic environments. SOAP used here can distinguish different radial and angular distributions of local atomic environments.

Interestingly, the energy calculated from DFT did not drop monotonically as the loss function did during the optimization. In fact, in phase I., an increase in energy can be observed. Such behavior comes from the non-linear relation between the energy and the local environment embedding. Therefore, it is possible to use LOSIKO to bypass some of the energy barriers during the conventional optimization process.

It is worthwhile to note that, in this case, the loss dropped to zero, meaning all local environments of atoms in the structure are driven to match exactly what's in the database. However, the atomic environments in real materials are complicated and most likely cannot be fully covered by the database. In these cases, the local environments in the database serve as an approximation of the real one. Considering the limited space of local structures for stable materials, such an approximation should provide enough accuracy.

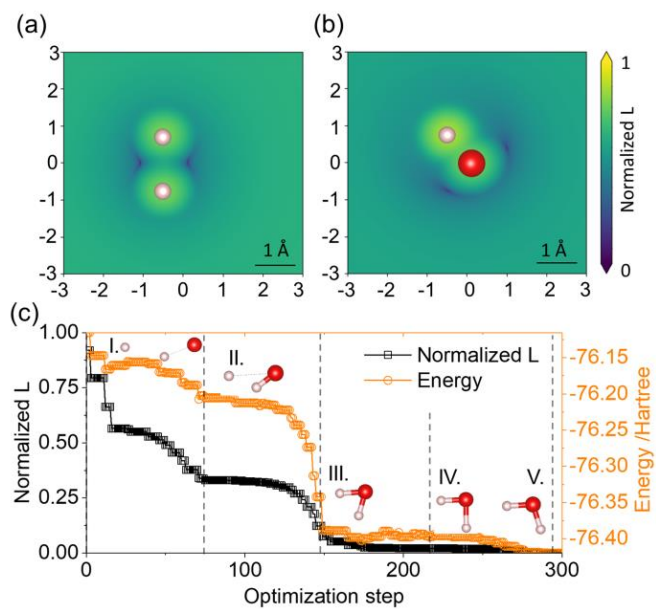

*Figure 3* Loss surfaces of an $H_2O$ molecule with (a) two hydrogen atoms fixed and (b) one oxygen and one hydrogen atoms fixed. (c) Loss function and the corresponding energy of the $H_2O$ molecule during the optimization. I., II., III., IV., and V. show the geometry of the $H_2O$ molecule at 0, 76, 148, 220, and 298 optimization steps, respectively.

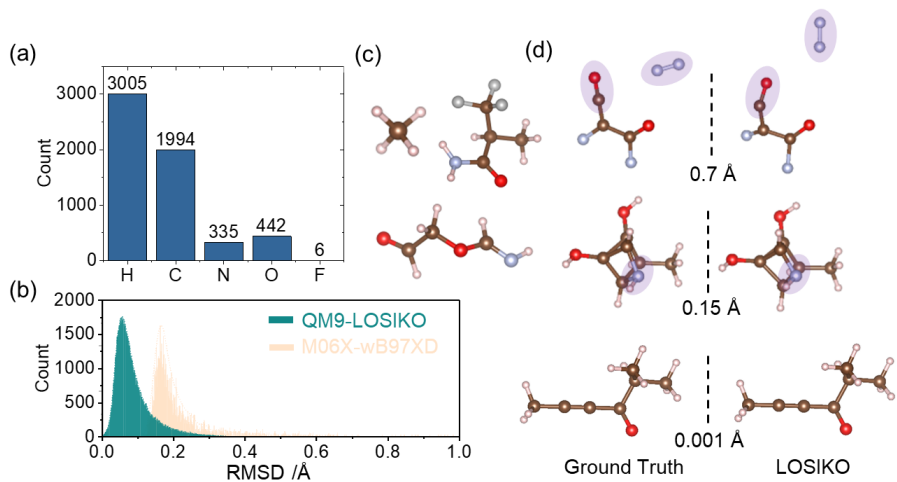

*Figure 4* *Optimization of QM9 structures using LOSIKO. (a) Element composition of the local environment embeddings in the database, (b) typical molecule structures used to construct the database, (c) RMSD distribution between the LOSIKO- and DFT-relaxed structures, (e) typical structures relaxed with LOSIKO and comparison with the ground truth.*

To quantify the performance of LOSIKO, we optimized the QM9 dataset which contains ~130,000 small molecules, using a fraction of the structures as the database. The database contains the local structure fingerprints of 3005 hydrogen, 1994 carbon, 335 nitrogen, 442 oxygen, and 6 fluorine atoms, as shown in **Figure 4(a)**. They are computed from 378 molecular structures sampled from the QM9 dataset. Typical structures in the database are shown in **Figure 4(c)**. The distribution of the root mean square displacement (RMSD) of the LOSIKO-relaxed structures from the ground truth is shown in **Figure 4(b)**. The mean displacement is on the order of ~0.08 Å. Such a value is smaller than the RMSD between structures relaxed using different DFT functionals. This further supports the finding that significant similarity exists between atomic environments in low-energy structures. Therefore, by choosing a representative database with low energy structures, one can achieve similar accuracy of quantum chemical approaches using LOSIKO. Apart from that, we found the deviation from the ground truth is smaller than some of the recently reported results using generative models, [27, 28] further showcasing its precision. However, we want to note that these generative models generate structures from a graph representation, which is more challenging than the task here.

It is worthwhile to note that LOSIKO leads to large RMSD of over 0.5Å in some cases. We checked the geometry of those structures and found that all of them contain molecular $N_2$ which is relatively far from the rest of the structure as shown in **Figure 4(d)**. Since isolated $N_2$ exists in the database, once LOSIKO matches the nitrogen atoms in the structure with the $N_2$, it will drive the $N_2$ away from the rest of the molecule to create an isolated $N_2$ environment. Some other structures with lower deviation are also shown in this figure. The deviation from the ground truth is determined by whether the local environment in the structure to relax is close to what's in the database. For example, as shown in the middle panel of **Figure 4(d)**, the nitrogen atom shows a relatively large displacement from the ground truth. This is due to the under representation of proper local environment of nitrogen atoms within compounds, reflected on a small number of nitrogen environments in the database. In contrast, some structures are very close to the ground truth with RMSD lower than 0.01 Å, as shown in the lower panel. This is because a very similar molecule (with only a shorter carbon chain) exists in the database.

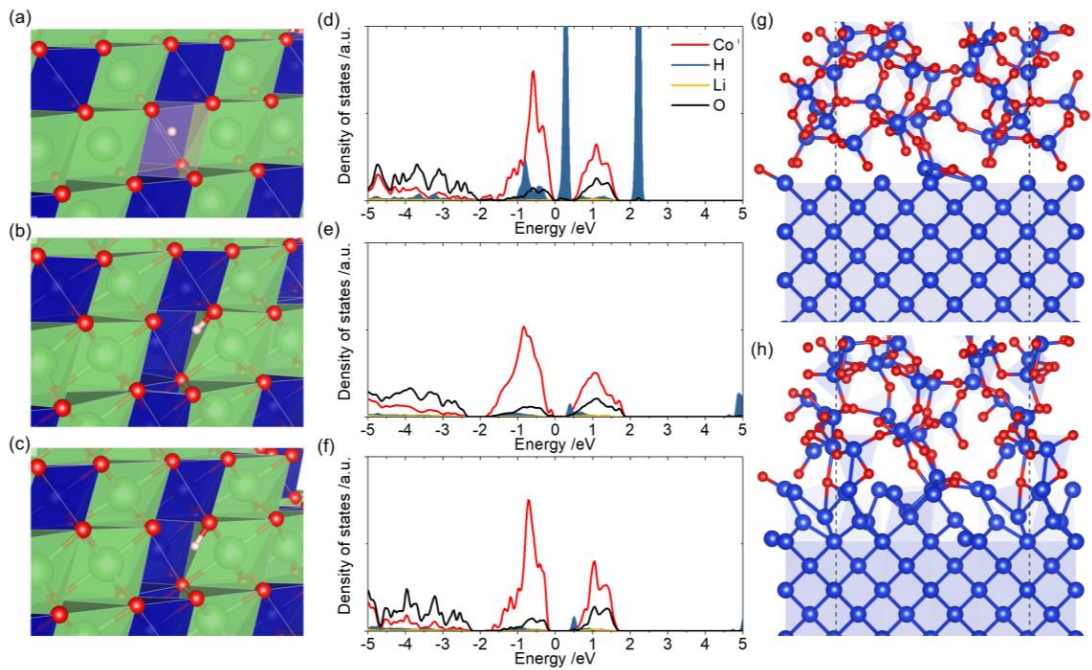

*Figure 5* Optimization of defect and interface structure using LOSIKO. (a) Unrelaxed, (b) LOSIKO-relaxed, and (c) DFT-relaxed proton defect in a $LiCoO_2$ crystal. (d) to (f) show the electronic density of states projected on different elements. The intensity of the hydrogen density of states is multiplied by a factor of 300 for visual clarity. (g) Unrelaxed and (h) LOSIKO-relaxed interface between crystalline silicon and amorphous $SiO_2$.

Obtaining atomic structures of complex systems remained a challenging task. This is particularly difficult in chemically diverse systems such as point defects and interfaces due to the large number of atoms in the system and the ill-defined bond topology. While quantum chemical methods can in principle resolve the topology issue, the complex systems are often too large to work with using these methods. We show that LOSIKO can be used to optimize these complex systems to reasonable structures. We first tested the proton defect in bulk $LiCoO_2$ crystal. During the processing of the cathode, water or other protic solvent molecules inevitably get in contact with the cathode particles and protons can exchange with the lithium ions in the material. [30, 31] The proton defect can affect the performance of the cathode. Here, we obtained the atomic coordinates of the proton defect by substituting a lithium atom with a hydrogen atom followed by a LOSIKO optimization, see **Figure 5(a)**. The database was constructed by assembly low energy structures (energy above hull < 200 meV atom$^{-1}$) in the Li-Co-O-H quaternary space from the Materials Project. The relaxed structure is shown in **Figure 5(b)**. The hydrogen atom is driven to bind with a nearby oxygen atom to form a hydroxyl group as opposed to an isolated proton in the unrelaxed structure. As shown in **Figure 5(c)**, the DFT relaxation also leads to a hydroxyl group, indicating such a configuration is energetically favored. In fact, we computed the energies of the unrelaxed, the LOSIKO-relaxed, and DFT-relaxed structures using DFT. The details of DFT computations are discussed in the Methodology section. The energies of the unrelaxed and the LOSIKO-relaxed structures are 11.04 eV and 0.62 eV higher than that of the DFT-relaxed structure, respectively. Such a result indicates that LOSIKO relaxed structures can serve as good approximations to the DFT local energy minimum. It is worthwhile to note that no structures explicitly contain a proton defect in the database, yet, LOSIKO can still relax the structure to the proximity of true energy minimum because the local environment of hydrogen in some hydroxides in the database such as LiOH, CoHO$_2$, and H$_2$O closely resembles that of the proton defect in LiCoO$_2$. To examine if the chemical nature of the defect has been captured, we computed the electronic density of states as a fingerprint. As shown in **Figure 5(d)** and **(e)**, the unrelaxed structure has a strong isolated defective state in the band gap of LiCoO$_2$ while such a state merged to combine with the oxygen p bands at the conduction band minimum in the LOSIKO-relaxed structure, indicating the strong O-H interaction. Such a feature of the electronic structure is in close resemblance to that of the DFT-relaxed one as shown in **Figure 5(f)**. Therefore, LOSIKO has driven the proton defect in LiCoO$_2$ to a chemically reasonable configuration.

Interfaces are more complicated to handle due to the chemical difference between the two sides of the interfaces. To test the performance of LOSIKO on such systems, we used the interface between crystalline silicon (c-Si) and amorphous silica (a-$SiO_2$) as an example. Such an interface widely exists in semiconductor devices. [32] Modeling such interfaces usually involves building large models followed by manual cleaning of the bonds and DFT relaxations, which are laborious and resource consuming. More importantly, current DFT methods need to compute the electronic structure of the entire structure to obtain the forces while only the region close to the interface core needs relaxation. This leads to significant resource waste and in extreme cases failure to deal with large systems. LOSIKO, on the other hand, can focus on the computation of the similarity measure for certain atoms in the whole system. Therefore, it has the advantage of being able to scale up to large systems. In this case, we build an interface by directly assembling a layer of c-Si and a layer of a-$SiO_2$. Both layers are directly cleaved from a bulk structure without extra treatment of the bonds. The idea is to use LOSIKO to drive the atoms at the interface to positions with reasonable bonding environments. We used bulk structures in the Si-O chemical space from the Materials Project to construct the database. Such a database contains not only the crystalline silicon structures but also tens of silicon oxides with different Si-to-O ratio, thus covering wide local environments for both silicon and oxygen. During the LOSIKO relaxation, only atoms that are within 5Å to the interfacial core are allowed to relax while other atoms are fixed. In addition, we added a gradually decaying random noise to the atom positions during the relaxation. This helps the structure to move out of the local minimum on the loss surface and achieve higher similarity to the templates in the database. The detail of adding such random noise is discussed in the Methodology section. The relaxation results are shown in **Figure 5(g)** and **(h)**. In comparison to the unrelaxed structure as shown in **Figure 5(g)**, more Si-O bonds are formed in the LOSIKO-relaxed model, leading to an energy drop of >5 eV Å$^2$. Interestingly, the surface of the crystalline silicon also experienced some reconfiguration, especially on the SiO2 layer. Such behavior has been observed both computationally and experimentally. [33-35] This case shows even with only bulk crystals as templates of local environments, one can still use LOSIKO to obtain reasonable structures of complex systems such as interfaces.

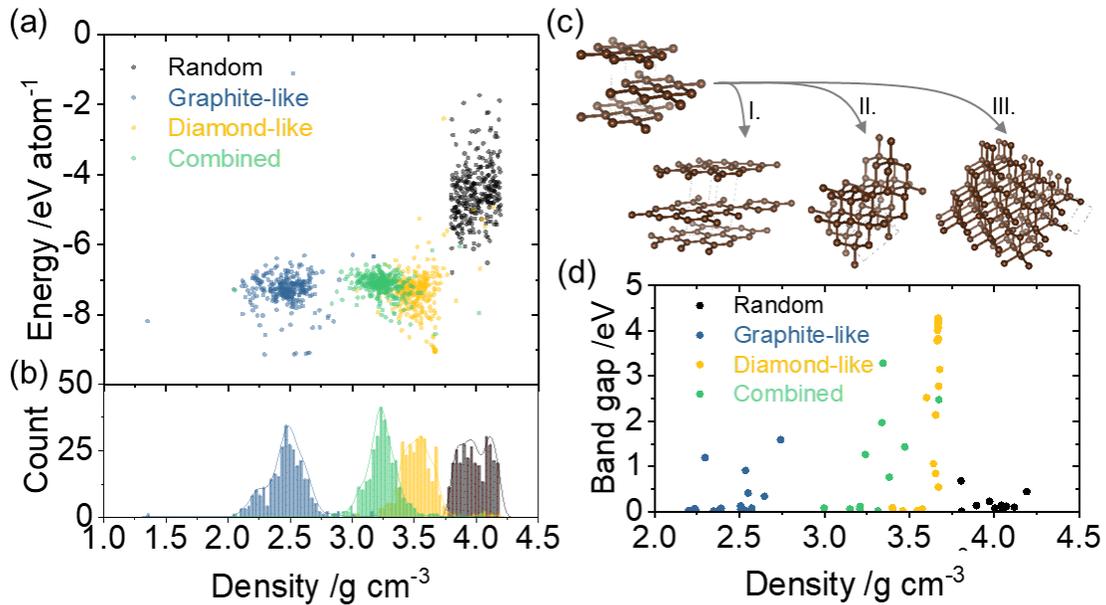

*Figure 6 Biased optimization and inverse design of materials with LOSIKO. The energy (a) and (b) density distribution of carbon structures that are randomly sampled and that are relaxed with LOSIKO with graphitic carbon environment, diamond carbon environment, and all possible carbon environments in low energy structures as database. (c) The transformation of a randomly sampled carbon structure to graphite (I.), diamond (II.), and distorted diamond (III.) with LOSIKO, respectively. (d) The electronic energy band gaps of the carbon structures sampled randomly and relaxed using LOSIKO but with different databases.*

Designing materials with desired properties, or inverse materials design, has been a grand challenge of materials science. [36, 37] Considering the property of a material is determined by its structure, one can, in principle, optimize the property by tuning its structural features.

LOSIKO provides a potential platform for such a task. By curating the database to only contain specific local structure fingerprints or tuning the weights in the loss function in Eq. (2) and eq. (8), the structural relaxation can be driven with bias. To test this, we used carbon materials as an example. For simplicity, we set all weights to 1 in the loss function as in Eq. (8) and use the database content to control the bias. We first generated 300 random structures with density ranging from 3.7 g cm$^{-3}$ to 4.2 g cm$^{-3}$ and then used these random structures as starting points for the LOSIKO optimization. To drive the structures with bias, we constructed three different databases. The first database only contains local environments of graphite. The relaxed structures using this database is denoted as *graphite-like*. The second database constitutes only the local environment in diamond and the relaxed structures will be named *diamond-like*. The last database includes multiple local environments from low-energy carbon structures from a previous ab initio random structure search. [20, 38] The energy threshold was set to 500 meV atom$^{-1}$ in this case. The relaxed structures using this dataset are denoted as *combined*. The results of the biased LOSIKO optimization are shown in **Figure 6 (a) and (b)**. The results of the biased LOSIKO optimization are shown in **Figure 6(a) and (b)**. The density of the graphite-like structures is the lowest among the three groups, with values ranging from 2.0 g cm$^{-3}$ to 2.7 g cm$^{-3}$. This result coincides with the density of graphitic carbon (~2.2 g cm$^{-3}$). This is expected because the structures are driven to be as close as possible to those six-member carbon rings with large interlayer spacing. The diamond-like structures are the densest among the three, with a mean value of ~3.55 g cm$^{-3}$. Such high density comes from the fully connected networks of dihedrals in the template. In-between are the structures relaxed with the *combined* database where many possible local structure fingerprints are included. It gives rise to the most diverse structures covering many diamond structures and several graphite-like ones. Energy-wise, all the LOSIKO-relaxed structures have much lower energy compared with the unrelaxed ones. Known carbon polymorphs including graphite, diamond, and some structures with penta-ring are found. [39] The one with the lowest energy is graphite, in agreement with previous theoretical findings. [39] In fact, we found several graphite structures with different stackings and interspacing in the search, which have similar energies and are ranked among the lowest. **Figure 6(c)** shows an example of how the same randomly sampled carbon structure is optimized to different final states with LOSIKO. The initial random structure features a layered configuration with 4 atoms in the unit cell. Each layer constitutes a square lattice and forms four-member carbon rings. Such a randomly generated structure is not close to any low energy structures. With different databases, the same structure can be optimized to graphite (see Path I. in **Figure 6(c)**), diamond (see Path II.), and distorted diamond (see Path III.). The optimized structures also display drastically different electronic properties. **Figure 6(d)** shows the non-zero electronic energy band gap of the relaxed structures. Among them, the *graphite-like* structure contains the lowest ratio of non-zero band gap materials (3.33%) while the *diamond-like* one contains the most (13.33%). This shows the capability of LOSIKO to drive the search of materials not only in the structural space but also the property space.

**Conclusions**

We develop a label-free data-driven LOcal SImilarity Kernel Optimization (LOSIKO) model to optimize atomic structures of materials in multiple phases, including isolated molecules, periodic crystals, and complex systems such as interfaces. It exploits the structural information encoded in embedding vectors, which are used to guide the atomic coordinates of the materials of interest by maximizing similarity measures between local environments of each atom to those stored in a database. Benchmark of such a method on the QM9 dataset containing ~130,000 molecules show remarkable performance in recovering molecular structures, with an RMSD of ~0.08 Å, smaller than that obtained from different DFT functionals. In addition, the method shows capabilities of relaxing complex systems to reasonable geometries without prior topological knowledge on bond connectivity. It captures the chemical feature of a protonic defect in a transition metal oxide which was validated by electronic structure calculation. It also builds an interface model with reasonable bonding environment for semiconductors. The success of the model further supports the idea that local atomic environments of stable molecules or materials only constitute a small portion of the entire structural space. In addition, we demonstrate that data can be used to drive the search of materials with target properties using LOSIKO. By pairing random structure searches and LOSIKO optimization with carefully curated databases, carbon structures with different local structural features are generated, leading to design capability of carbon materials with different densities and electronic band gaps.

**Methodology**

*Local atomic environment encoding.* The local environment $\chi_i$ of an atom $i$ is defined as the Cartesian coordinates of its surrounding atoms $j$ within a cutoff $r_c$:

$$\chi_i = \chi_i(\boldsymbol{r}_j) \quad \text{with} \quad |\boldsymbol{r}_i - \boldsymbol{r}_j| \leq r_c \tag{3}$$

Further embedding is needed for a description that is invariant to translation, rotation, and permutation operations. In general, there are two ways to embed such local environments.

One way is to use physics-inspired definitions to directly convert the local environments to vectors. So far, many embeddings (or more frequently known as representations) have been developed, the most well-known being smooth overlap of atomic positions (SOAP) and atom-centered symmetry functions (ACSF). [21, 40, 41] The other way is to use machine-learned embeddings. [42] Such embeddings are usually the output of a layer in the neural net which sits before the fully connected multilayer perceptron layers. Different model architectures can be used for the neural nets as long as the symmetry invariance is fulfilled. In this work, we adopt the SOAP formalism due to its effectiveness and simplicity. In principle, other embeddings can also be used. SOAP encodes the local atomic environment $\rho_i$ of atom $i$ by placing gaussian functions on each surrounding atom followed by expanding it into radial functions $g(|\mathbf{r}|)$ and spherical harmonics $Y_l m(\hat{r})$: [21]

$$\rho_i(\mathbf{r}_j) = \sum_{|\mathbf{r}_i - \mathbf{r}_j| \leq r_c} exp\left(-\frac{|\mathbf{r}_i - \mathbf{r}_j|^2}{2\sigma^2}\right)$$

$$= \sum_{nlm} c^\alpha_{nlm} g_n(|r|) Y_l m(\hat{r}) \qquad (4)$$

where $r_c$ is the cutoff radius for characterizing the local environments and $\sigma$ is the width of the gaussian functions. The power spectrum is then taken as the final embedding by computing on the expansion coefficients up to certain angular and radial indices, i.e., $l_{max}$ and $n_{max}$, respectively

$$k^\alpha_{nn'l} = \sqrt{\frac{8}{2l+1}} \sum_m (c^\alpha_{nlm})^* c^\alpha_{n'lm} \qquad (5)$$

In this work, the DScribe package is used to carry out such conversion. [43] The choice of parameters $l_{max}, n_{max}, r_c$, and $\sigma$ depends on the system of interest and the values used in this work are listed in **Table S1**.

*Similarity kernels.* Similarities between atomic environments can be computed in several ways. In the original definition, the SOAP kernel is computed as the normalized polynomial kernel of the power spectrums while other kernel functions can be used as well:

$$K(p_i, p_j) = \left(\frac{p_i \cdot p_j}{\sqrt{p_i \cdot p_i \, p_j \cdot p_j}}\right)^d \tag{6}$$

where $p_i = \{k^{\alpha}_{nn'l}\}_i$ is the vector constructed from the expansion coefficients of atom $i$ and $d$ is the kernel degree. Usually, such kernels are converted to distance metrics to enable easy computation of the loss function. In this work, we use the $L_2$ distance between normalized $p$'s as a measure of similarity between local atomic environments [44]

$$D(p_i, p_j) = (p_i - p_j) \cdot (p_i - p_j) \tag{7}$$

Despite that such a $D$ is not a kernel function itself, it provides a simple yet effective means to access the similarity between different atomic environments.

*Optimization.* The optimization of the atomic structures is carried out by minimizing the loss function through moving the positions of each atom in the structure. A detailed computational flowchart is shown in **Figure S2**. The loss function used in LOSIKO is defined as the weighted distances between the atomic environments to the targets $p_i^t$ in the database

$$\hat{L}(\mathbf{r}_i) = \sum_{i=0}^{n} c_i D(p_i, p_i^t)) \tag{8}$$

which is equivalent to maximizing the sum of similarity kernels over all $n$ atoms as in Eq. (2). The coefficients determine the bias towards a local environment during the optimization. For simplicity, we set all the weights to 1 in this work. $\hat{L}$ values zero when all local environments are exactly the same with the target in the database while a larger value means a larger structural deviation. To minimize the loss function $\hat{L}$, we used the Nelder-Mead method as implemented in SciPy. [44, 45] For a typical LOSIKO optimization, we first construct a database of the local environments by computing the local environment embeddings for each atom in structures from the database. Then, the atomic environment embeddings are computed for the structure to relax, followed by searching for their closest targets in the database using the distance function $D$ or the similarity kernel $K$. Finally, the atomic positions (and the lattice vectors for crystal systems) are relaxed to minimize the loss function as defined above in Eq. (8). It is worthwhile to note that such optimization is sometimes hindered by local barriers. Therefore, we introduced a gradually decaying random noise to the atomic positions during the relaxation

$$\boldsymbol{r}_{iter} \leftarrow \boldsymbol{r}_{iter} + N(0, \sigma_{iter}^2) \tag{9}$$

where $N(0, \sigma_{iter}^2)$ is a Gaussian distribution with a decaying standard deviation $\sigma_{iter}^2 = \sigma_0^2 \exp(-\alpha\, iter)$. $\sigma_0^2$ is the pre-exponential factor and $\alpha$ is a coefficient defining how fast the noise decays to zero.

*Databases.* The structures used to construct databases are collected from sources including the Materials Project, the QM9 database, as well as previously published random structure search results. [18-20, 23, 46] For the case of relaxation on small molecules from QM9, the database was built upon the first 100 structures within QM9 which contain simple molecules like $CH_4$ and $H_2O$, and a uniform sampling of the rest structures with a stride of 500. In total, 378 molecules were used to construct the database, constituting a sparse sampling of the 0.2% of the entire QM9 databank. During optimization, these structures were removed from the database. For the case of proton defect in $LiCoO_2$ and $Si/SiO_2$ interfaces, the databases were constructed from structures in the Materials Project. All possible phases are considered in the chemical systems of interest and a filter was applied to screen out any structure with an energy above hull over 0.3 eV atom$^{-1}$. For the case of biased carbon search, the database was constructed from a carbon databank generated using random structure searching. Similarly, an energy filter was applied to eliminate any structures with an energy 0.5 eV atoms$^{-1}$ higher than the lowest entry.

*Quantum Chemical Computations.* All quantum chemical computations on isolated molecules are carried out using the DFT module implemented in the Gaussian 16 package.[47] Three hybrid functionals, B3LYP, M062X and wB97XD are employed. The 6-31G(d) basis set is used and the convergence criterion for root mean square deviations in electron density matrix elements in the SCF algorithm is set to $10^{-8}$. For the periodic systems, first-principles calculations are carried out on the level of DFT using the VASP6.3 package with the PBE exchange–correlation functional.[48] A planewave basis with a cutoff energy of 520 eV together with Monkhorst-Pack grids with a spacing of 0.04 Å$^{-1}$ is adopted. The convergence for electron self-consistent computations and structural optimizations are set to $10^{-5}$ eV atom$^{-1}$ and $10^{-3}$ eV atom$^{-1}$, respectively.

**Data Availability**

The code of LOSIKO will be uploaded to and maintained at the GitHub page (https://github.com/luzihen/LOSIKO). All the structural data as well as the DFT calculation results will be uploaded to the same repository.

**Acknowledgements**

Z.L. thanks Dr. Jiapeng Liu and Dr. Zhaofu Zhang for discussions and proofreading the manuscript.# References

[1]   A. R. Oganov, C. J. Pickard, Q. Zhu, R. J. Needs, *Nat. Rev. Mater.* 2019, 4, 331.
[2]   S. M. Woodley, R. Catlow, *Nat. Mater.* 2008, 7, 937.
[3]   X. M. He, D. C. Carter, *Nature* 1992, 358, 209.
[4]   D.-H. Kwon, K. M. Kim, J. H. Jang, J. M. Jeon, M. H. Lee, G. H. Kim, X.-S. Li, G.-S. Park, B. Lee, S. Han, *Nat. Nanotechnol.* 2010, 5, 148.
[5]   S. Pennycook, *Annual review of materials science* 1992, 22, 171.
[6]   N. Pavliček, L. Gross, *Nature Reviews Chemistry* 2017, 1, 1.
[7]   R. F. Ziesche, N. Kardjilov, W. Kockelmann, D. J. Brett, P. R. Shearing, *Joule* 2022, 6, 35.
[8]   P. Qi, J. Wang, X. Djitcheu, D. He, H. Liu, Q. Zhang, *RSC Adv.* 2022, 12, 1216.
[9]   Z. Lu, *Materials Reports: Energy* 2021, 1, 100047.
[10]  I. Errea, F. Belli, L. Monacelli, A. Sanna, T. Koretsune, T. Tadano, R. Bianco, M. Calandra, R. Arita, F. Mauri, *Nature* 2020, 578, 66.
[11]  J. K. Nørskov, T. Bligaard, J. Rossmeisl, C. H. Christensen, *Nature chemistry* 2009, 1, 37.
[12]  C. Oostenbrink, A. Villa, A. E. Mark, W. F. Van Gunsteren, *Journal of computational chemistry* 2004, 25, 1656.
[13]  J. Behler, M. Parrinello, *Phys. Rev. Lett.* 2007, 98, 146401.
[14]  A. P. Bartók, M. C. Payne, R. Kondor, G. Csányi, *Phys. Rev. Lett.* 2010, 104, 136403.
[15]  O. T. Unke, S. Chmiela, H. E. Sauceda, M. Gastegger, I. Poltavsky, K. T. Schütt, A. Tkatchenko, K.-R. Müller, *Chem. Rev.* 2021, 121, 10142.
[16]  L. Wang, T. Maxisch, G. Ceder, *Phys. Rev. B* 2006, 73, 195107.
[17]  A. Jain, G. Hautier, S. P. Ong, C. J. Moore, C. C. Fischer, K. A. Persson, G. Ceder, *Phys. Rev. B* 2011, 84, 045115.
[18]  R. Ramakrishnan, P. O. Dral, M. Rupp, O. A. Von Lilienfeld, *Scientific data* 2014, 1, 1.
[19]  C. Pickard, *Materiials Cloud Archive* 2020, 10.
[20]  M. Martinez-Canales, C. J. Pickard, R. J. Needs, *Phys. Rev. Lett.* 2012, 108, 045704.
[21]  A. P. Bartók, R. Kondor, G. Csányi, *Phys. Rev. B* 2013, 87, 184115.
[22]  B. Cheng, R.-R. Griffiths, S. Wengert, C. Kunkel, T. Stenczel, B. Zhu, V. L. Deringer, N. Bernstein, J. T. Margraf, K. Reuter, *Accounts of Chemical Research* 2020, 53, 1981.
[23]  A. Jain, S. P. Ong, G. Hautier, W. Chen, W. D. Richards, S. Dacek, S. Cholia, D. Gunter, D. Skinner, G. Ceder, *APL materials* 2013, 1, 011002.
24

Supplementary Information

Accessing and driving atomic structures of molecules, crystals, and complex systems with local similarity kernels

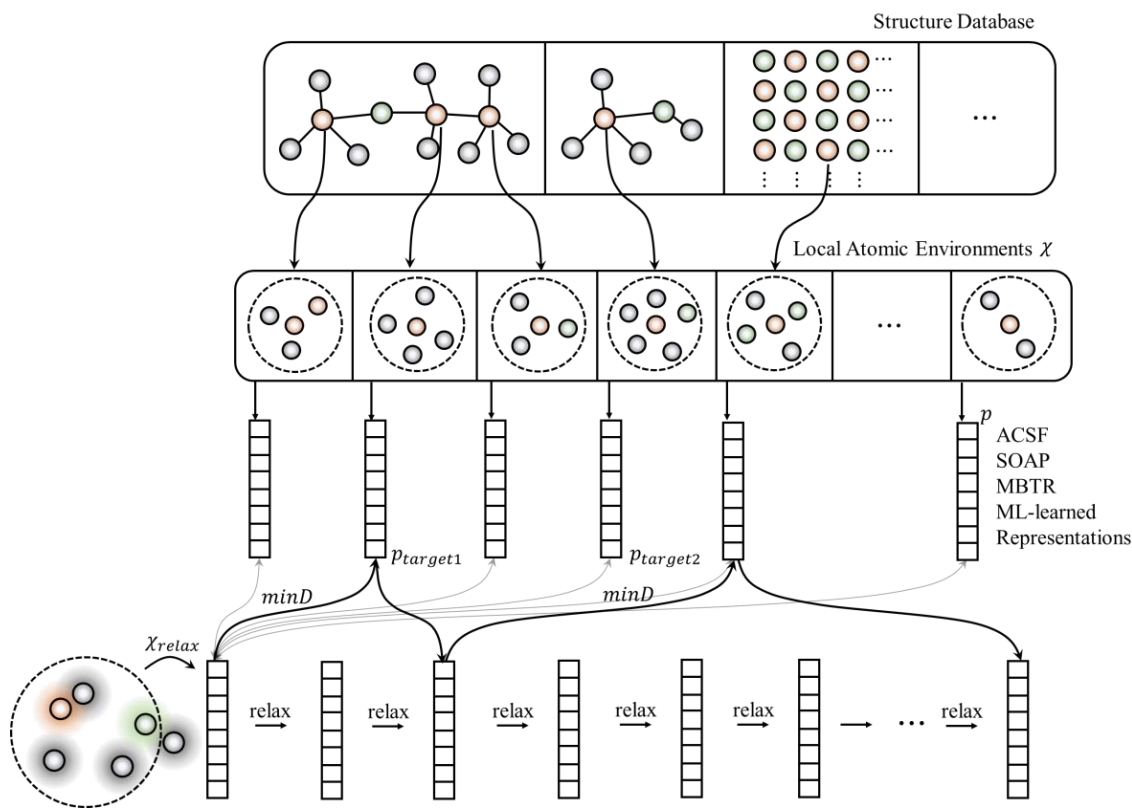

**Figure S1** Illustration of the LOSIKO optimization process.

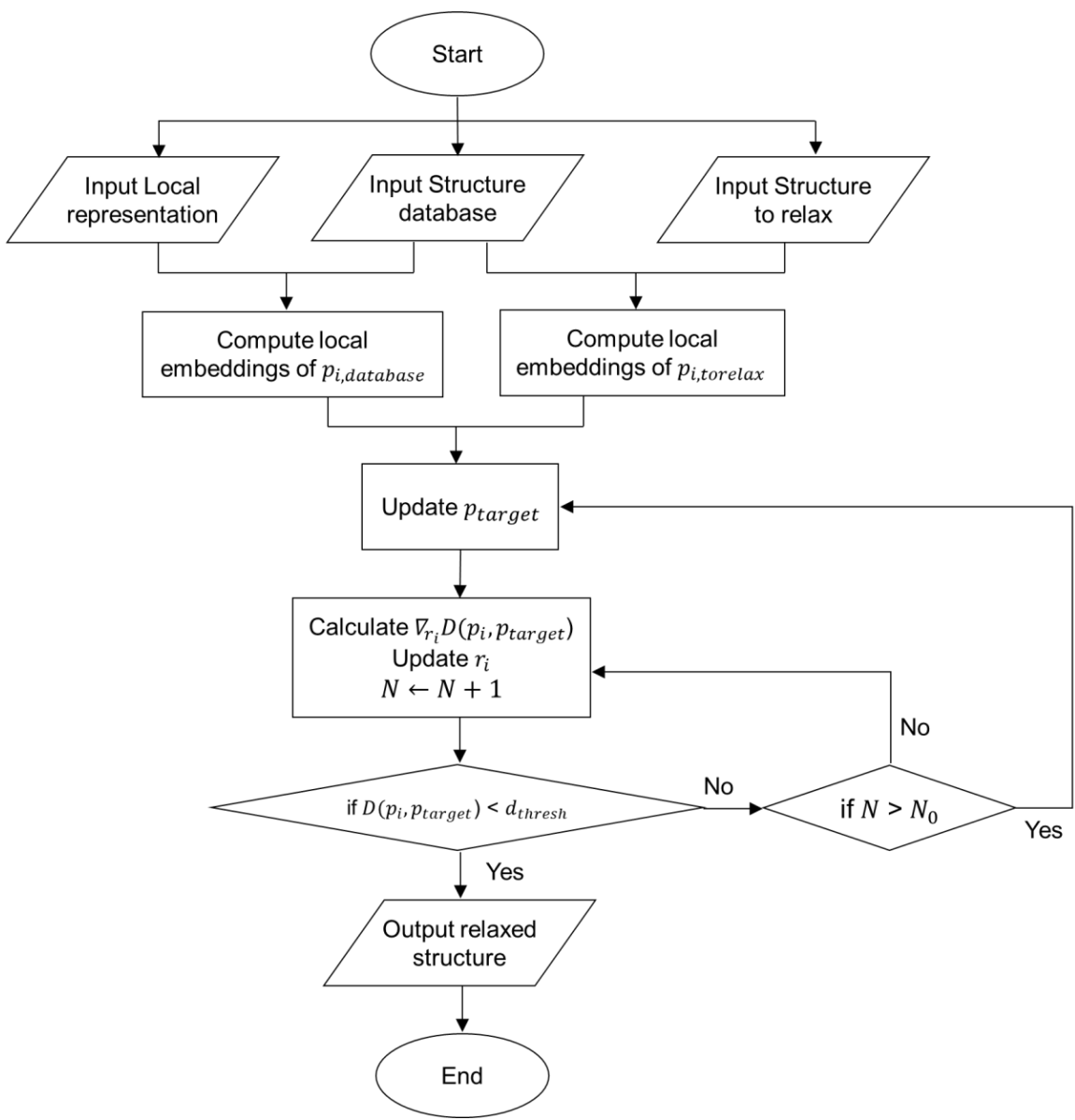

**Figure S2** Computation flowchart of LOSIKO

**Table S1** Parameters used for the local environment embedding and the LOSIKO optimization

| Case | Database | Local environment embedding | Parameters |
|---|---|---|---|
| $H_2O$ | Relaxed $H_2O$ | SOAP | $r_{cut} = 4$, $n_{max} = 4$, $l_{max} = 4$, $\sigma = 0.25$, $\sigma_0^2 = 0$ |
| Small molecules QM9 | 378 structures sampled from QM9 | SOAP | $r_{cut} = 4$, $n_{max} = 6$, $l_{max} = 6$, $\sigma = 0.5$, $\sigma_0^2 = 0$ |
| Proton defect in $LiCoO_2$ | Materials Project | SOAP | $r_{cut} = 4$, $n_{max} = 4$, $l_{max} = 2$, $\sigma = 0.7$, $\sigma_0^2 = 0$ |
| c-Si/a-$SiO_2$ interface | Materials Project | SOAP | $r_{cut} = 4$, $n_{max} = 8$, $l_{max} = 6$, $\sigma = 0.25$, $\sigma_0^2 = 0.2$, $\alpha = 0.05$ |
| carbon | Random structure search results | SOAP | $r_{cut} = 2.3$, $n_{max} = 8$, $l_{max} = 6$, $\sigma = 0.25$, $\sigma_0^2 = 0$ |